\documentclass[journal,10pt,doublecolumn,singlespace]{IEEEtran}

\usepackage{cite,graphicx,subfig,multirow,array}

\usepackage{amsfonts,amssymb}
\usepackage[cmex10]{amsmath}
\usepackage{algorithmic}
\usepackage[ruled]{algorithm2e}	
\usepackage{epstopdf}
\usepackage{hyperref}
\usepackage{verbatim}

\usepackage[usenames,dvipsnames]{color}
\usepackage{soul}

\usepackage{colortbl}
\definecolor{mygray}{gray}{.9}

\DeclareMathOperator*{\argmax}{argmax}

\begin{document} 
	

\title{Attention-Guided Progressive Neural Texture Fusion for High Dynamic Range Image Restoration}

	
\author{Jie Chen, Zaifeng~Yang, Tsz Nam Chan, Hui Li, Junhui Hou, and Lap-Pui Chau\\

\thanks{J. Chen and T. N. Chan are with the Department of Computer Science, Hong Kong Baptist University (e-mail: chenjie@comp.hkbu.edu.hk and edisonchan@comp.hkbu.edu.hk).}
\thanks{Z. Yang is with the Institute of High Performance Computing, A*STAR, Singapore (e-mail: yang\_zaifeng@ihpc.a-star.edu.sg).}
\thanks{H. Li is with the Imaging Algorithm Research Department, vivo Mobile Communication Co. Ltd, Shenzhen, China. (e-mail: lihui@vivo.com)}
\thanks{J. Hou is with the Department of Computer Science, City University of Hong Kong (e-mail: jh.hou@cityu.edu.hk).}
\thanks{L.-P. Chau is with the School of Electrical \& Electronic Engineering, Nanyang Technological University, Singapore (e-mail: elpchau@ntu.edu.sg).}

}

\markboth{}
{\MakeLowercase{\textit{Chen et al.}}:}
	
\maketitle

\begin{abstract}

High Dynamic Range (HDR) imaging via multi-exposure fusion is an important task for most modern imaging platforms. In spite of recent developments in both hardware and algorithm innovations, challenges remain over content association ambiguities caused by saturation, motion, and various artifacts introduced during multi-exposure fusion such as ghosting, noise, and blur. In this work, we propose an Attention-guided Progressive Neural Texture Fusion (APNT-Fusion) HDR restoration model which aims to address these issues within one framework. An efficient two-stream structure is proposed which separately focuses on texture feature transfer over saturated regions and multi-exposure tonal and texture feature fusion. A neural feature transfer mechanism is proposed which establishes spatial correspondence between different exposures based on multi-scale VGG features in the masked saturated HDR domain for discriminative contextual clues over the ambiguous image areas. A progressive texture blending module is designed to blend the encoded two-stream features in a multi-scale and progressive manner. In addition, we introduce several novel attention mechanisms, i.e., the motion attention module detects and suppresses the content discrepancies among the reference images; the saturation attention module facilitates differentiating the misalignment caused by saturation from those caused by motion; and the scale attention module ensures texture blending consistency between different coder/decoder scales. We carry out comprehensive qualitative and quantitative evaluations and ablation studies, which validate that these novel modules work coherently under the same framework and outperform state-of-the-art methods.

\end{abstract}
\begin{keywords}
High dynamic range imaging, neural feature Transfer, multi-scale fusion, visual attention
\end{keywords}

\section{Introduction}

\IEEEPARstart{T}{he} intensity of light rays vary in a great range in natural scenes. Under a common outdoor scenario, the luminance 
variation covers the range of $10^5~\text{cd}/\text{m}^2$. After millions of years of evolution, the human iris and brain are able to constantly adapt and adjust the responses to such strong stimulant variations, and perceive through the bright and dark contents of the scene. 
Most camera sensors, however, only cover a dynamic range of around $10^3~\text{cd}/\text{m}^2$, which makes single image capture prone to show either over-exposed or contrast-constrained, noise-inflicted pixels. 



To achieve High Dynamic Range (HDR) imaging, there are two practical strategies. 
The first strategy is to work in the radiance domain. 
By designing the Camera Response Functions (CRF), the sensor sensitivity for certain luminance ranges can be compressed so that a broader dynamic range can be covered \cite{li2018clustering}; however, this strategy sacrifices the imaging quality for potential target intensity ranges. Dedicated optical systems have been designed to capture HDR snapshots directly \cite{nayar2000high, tumblin2005want, serrano2016convolutional}. These systems are generally robust against camera and scene motion; however they are too bulky and expensive to be accepted by the consumer markets. In addition, with the development of semiconductor manufacturing technologies, pixel-level sensor structures can now be designed in which sensing area under the same color filter unit is split into pixels with different exposure settings \cite{ovt4c}. Though these sensor systems alleviate the alignment issues between the LDR pixels, the image resolution is traded-off for higher dynamic range. Additionally, the differences in exposure settings cause additional challenges (e.g., longer exposures introduce motion blur, while shorter ones are subject to strong sensor noise), which require advanced fusion algorithms to address.

The second strategy is to work in the image domain, i.e., via fusing a sequence of differently exposed (Low Dynamic Range) LDR exposures \cite{ma2019deep}.
The challenges for satisfactory fusion lie in the content association ambiguities caused by saturation, and the motion from both camera and dynamic objects.
Both active measures such as flow-based pixel alignment \cite{kalantari2017deep}, and preventive measures such as attention masking \cite{yan2019attention} and patch-based decomposition \cite{yan2019attention} have been investigated to tackle these challenges. Active measures strive to align pixels displaced by camera and object motion; however, they cannot handle regions with correspondence ambiguities from occlusion, non-rigid transformation, and specifically, saturation. These ambiguities result in warping artifacts caused by wrong correspondence predictions. 
Preventive measures passively avoid incorporating differing textures from fusion. Attention to motion helps to avoid ghosting artifacts, but it also limits useful information to be transferred from well-exposed references. Fusing artifacts such as halo \cite{mertens2009exposure} and blur also occur often.

In conclusion, the limitations and challenges for existing fusion methods can be summarized into the following three aspects: \textit{\ul{first}}, how to differentiate misalignment ambiguities caused by saturation and motion, and subsequently adopt different strategies for the two, i.e., texture transfer and signal regularization; \textit{\ul{second}}, how to accurately locate reference information over the ambiguities caused by saturation--especially when the saturation area is large and when it is overlapped with motion; and  \textit{\ul{finally}}, how to fully explore the characteristic information from different captures for general vision enhancement goals such as noise reduction and avoidance of common fusion artifacts, e.g., halo and blur.


In this work, we propose an \textit{Attention-guided Progressive Neural Texture Fusion} (APNT-Fusion) framework for HDR restoration which addresses the challenges of motion-induced ghosting artifacts prevention and texture transfer over saturated regions efficiently within one framework. 
Both qualitative and quantitative evaluations validate the advantages of our method against existing solutions. The novelty and technical contributions of this work can be generalized as follows:
\begin{itemize}
\item we propose an efficient \textit{two-stream} structure which separately focuses on texture feature transfer over saturated regions and fusion of motion suppressed multi-exposure features. A Progressive Texture Blending (PTB) module is designed to blend the encoded features in a multi-scale and progressive manner, and produces the final restoration results;
\item we propose a novel and efficient Neural Feature Transfer (NFT) mechanism which establishes spatial correspondence between different exposures based on multi-scale VGG features in the Masked Saturated HDR (\textit{MS-HDR}) domain. This mechanism provides discriminative contextual clues over the ambiguous image areas--especially over large saturation areas overlapped with motion--and provides accurate texture reference;
\item we introduce several novel attention mechanisms, i.e., the Motion Attention module detects and suppresses the content discrepancies among the reference images; the Saturation Attention module facilitates differentiating the misalignment caused by saturation from those caused by motion--therefore encouraging texture transfer to regions with missing contents; and the Scale Attention module ensures texture blending consistency between different decoder scales. These attention mechanisms cooperate well under the same framework and greatly improve the restoration performance as to be validated in our ablation studies.
\end{itemize}

The rest of the paper is organized as follows: Sec. \ref{sec:relatedworks} introduces related works, Sec. \ref{sec:method} explains the details of the proposed method, Sec. \ref{sec:evaluation} comprehensively evaluates the proposed model and compares with existing methods, ablation studies are carried out in Sec. \ref{sec:ablation_study}, and Sec. \ref{sec:conclusion} concludes the paper.

\section{Related Works} \label{sec:relatedworks}

Suppose we have a sequence of 3 differently exposed images of the same target scene $\mathcal{I}=\{I_s,~I_m,~I_l\}$, where subscripts $s$, $m$, and $l$ stand for short, medium, and long exposures, respectively. The most straightforward operation to produce a fused HDR is via pixel weighted summation:
\begin{equation}\label{eqn_hdrcombine}
\sum_{k=l,m,s}{\omega_k\cdot \mathcal{E}(I_k)},
\end{equation}
where $\omega_k$ is the weight related to each pixel’s intensity value and sensor noise characteristics \cite{kalantari2017deep}. $\mathcal{E}$ stands for an operator that brings all images to the same exposure energy level. Eq. (\ref{eqn_hdrcombine}) assumes all pixels are perfectly aligned between the images, which is generally not true due to various factors such as camera motion and dynamic scene contents. Numerous methods have been proposed over recent years to alleviate the ghosting artifacts, and these methods can be generalized into the following categories.

\subsection{Pixel Rejection Methods}
One direct way to reduce ghosting artifacts is to choose one image as reference, and detect motion areas between the reference and non-reference images, and exclude these pixels during fusion. Usually, the medium exposure image is chosen as a reference. The problem of ghost detection is similar to those of motion detection, with the added challenge that scene contents could be visually different under different exposure settings. To counter such a challenge, gradient maps \cite{zhang2019gradient} and the median threshold bitmaps \cite{pece2010bitmap} have been used for inconsistency detection. Efforts are seen in using mathematical models to optimize a correct ghost map \cite{granados2013automatic}. Rank minimization techniques have also been investigated by Lee et al. \cite{lee2014ghost} to ensure a high quality fusion. By rejecting the ghosting pixels, these methods lost the valuable information from these areas at the same time. Ma et al. \cite{ma2017robust} proposed a structural patch decomposition approach which decomposes image patches into three components: strength, structure, and mean intensity. The three patch components are processed separately and then fused with good ghost removal effect. Li et al. further enhanced this structural patch decomposition approach by reducing halo \cite{li2020fast} and preserving edges \cite{li2021detail}.

\subsection{Content Association and Registration Methods}
Another line of work aims at aligning the pixels before HDR fusion. Kang et al. \cite{kang2003high} register the pixels between video frames using optical flow \cite{lucas1981iterative} and merge the associated pixels to reduce artifacts. Jinno and Okuda \cite{jinno2008motion} estimate the pixel displacements, occlusion, and saturated regions with a Markov random field model. Oh et al. \cite{oh2015robust} simultaneously align LDR images and detect outliers that break the rank-1 structure of LDR images for robust HDR fusion. Precise association of pixels between instances with large motion is a challenging problem by itself, and unavoidable alignment artifacts are difficult to avoid in a pixel-level framework.

\begin{figure*}[t]
	\centering
	\includegraphics[width= 1\linewidth]{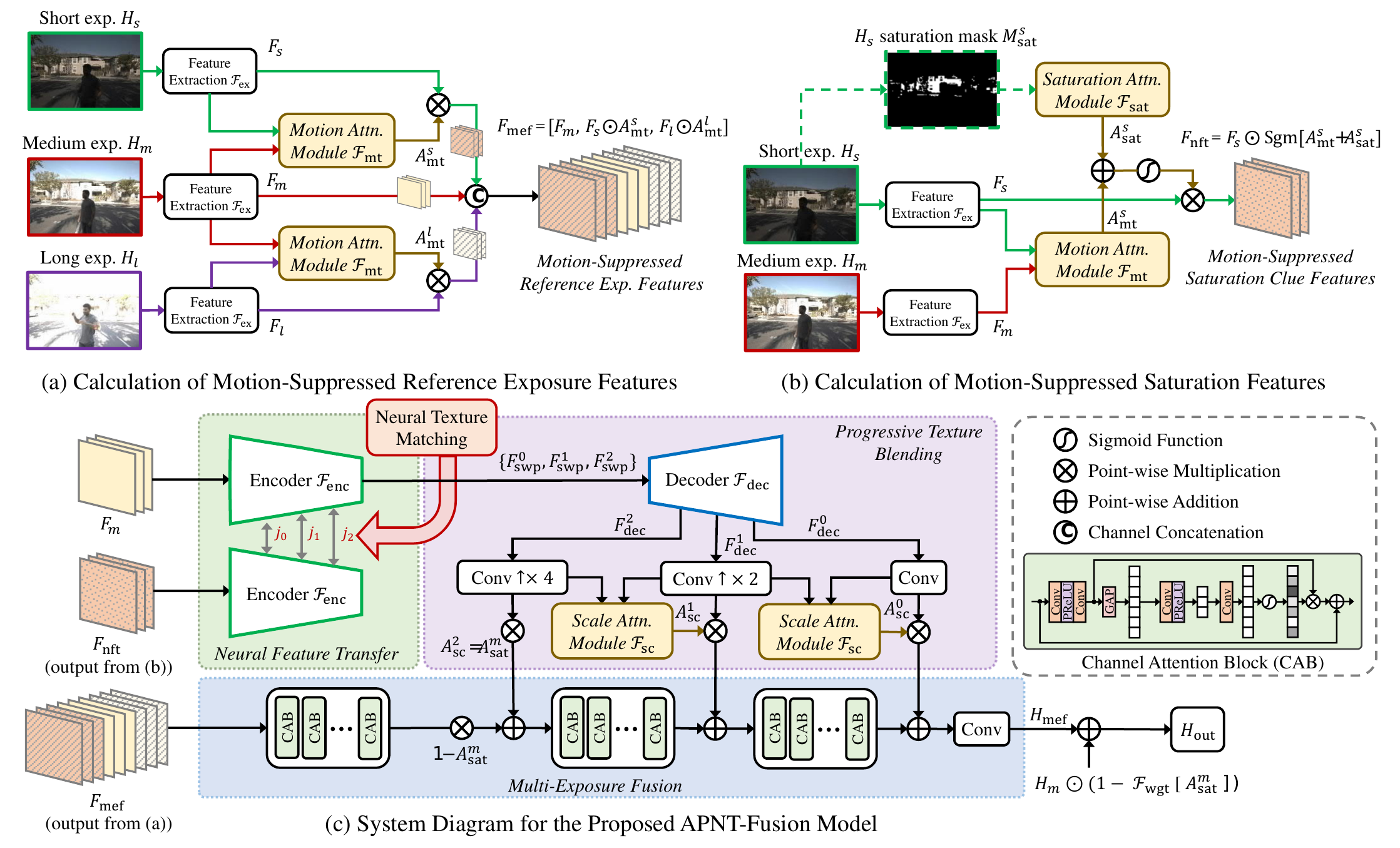}
	\caption{The system diagram of our proposed network. (a) and (b) give the details for the calculation of the \textit{Motion-Suppressed Reference Exposure Features} and the \textit{Motion-Suppressed Saturation Clue Features}, respectively. (c) gives the overall system structure with features from (a) and (b) as inputs. The detail for the Neural Texture Matching module can be found in Fig. \ref{fig:matchfeature}.}
	\label{fig:systemDiagram}
\end{figure*}

\subsection{DNN based methods}
Deep Neural Networks show their advantages in a wide range of computational imaging and image restoration problems  \cite{chen2018light, zhang2017beyond}. Wu et al. \cite{Wu2018deep} formulated HDR imaging as an image translation problem. Missing contents caused by occlusion, over-/under-exposure are hallucinated. Eilertsen et al. \cite{eilertsen2017hdr} proposed to predict an HDR image based on a single LDR input with an autoencoder structure. Endo et al. \cite{endo2017deep} achieved the same target by combining multiple intermediate LDR predictions from a single LDR using DNNs. For these methods, since details are added based on knowledge learned from distributions of other images in the training dataset, the predictions might be incorrect for a specific image. 
Kalantari et al. \cite{kalantari2017deep} use DNNs to merge and refine the LDR images based on pre-aligned image tensors with optical flow. Besides the possible alignment error, this popular method is limited due to its constrained mapping space. Yan et al. \cite{yan2019attention} proposed to guide the merging of LDR images via an attention model over the reference image. Deng et al. \cite{deng2021deep} proposed a deep coupled feedback network to achieve multiple exposure fusion and super-resolution simultaneously. Attention has shown to be an extremely useful tool for computer vision problems which boosts the robustness of the network by allowing models to focus on only the relevant information. However, when attention maps are used to highlight reference content inconsistency, they suppress ghosting artifacts at the expense of useful texture from being transferred to saturated regions. 

In general, current state-of-the-art solutions show satisfactory performances in avoiding ghosts after LDR fusion. The performance of transferring textures and colors from ambiguous regions with motion is limited. Fusion quality issues such as color fidelity and signal noisiness--which are the main challenges for mobile imaging platforms--have not been addressed systematically in one framework.



\begin{figure*}[t]
	\centering
	\includegraphics[width=0.96\linewidth]{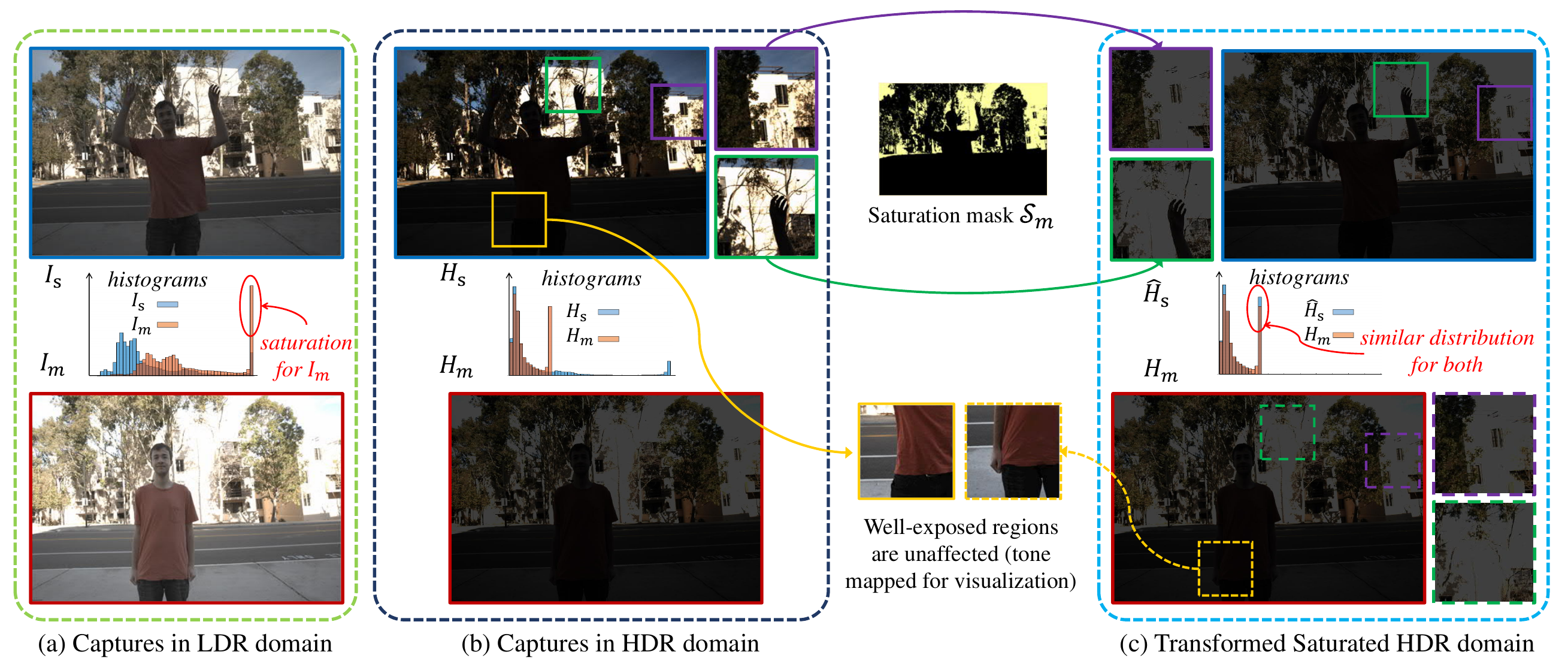}
	\caption{Exposure domain transform from LDR domain to Saturated HDR (S-HDR) domain. Histograms for both $I_s$ and $I_m$ captures have been plotted to highlight the dynamic range change. The histograms of both captures in S-HDR domain show very similar distributions.}
	\label{fig:domainTansform}
\end{figure*}

\section{Proposed Method} \label{sec:method}


Given a multi-exposure LDR image sequence $\mathcal{I}=\{I_k | I_k\in\mathbb{R}^{H\times W\times C}\}_{k=l,m,s}$, our model $\mathcal{N}$ aims at restoring a well-exposed HDR image $H_\text{out}\in\mathbb{R}^{H\times W\times C}$, whose contents are accurately aligned to the medium exposed image $I_m$, with over-saturated pixels in $I_m$ compensated by references from $I_s$, and under-exposed regions regularized by $I_l$: 
\begin{equation}
H_\text{out}= \mathcal{N}(I_s, I_m, I_l;~\theta),
\end{equation}
here $\theta$ is the set of model parameters to be learned. $H$ and $W$ indicate the spatial resolution, and $C$ indicates the image channel number, respectively.

The system diagram of our proposed \textit{Attention-guided Progressive Neural Texture Fusion} (APNT-Fusion) HDR restoration framework is shown in Fig. \ref{fig:systemDiagram}(c). The system consists of three main sub-modules: 
\begin{itemize}
	\item the \textit{Multi-Exposure Fusion} (MEF) module fuses signals from different exposure levels and maps them to an optimal regularized signal subspace;
	\item the \textit{Neural Feature Transfer} (NFT) module establishes the spatial correspondence between different images based on encoded VGG features in the Masked Saturated HDR (\textit{MS-HDR}) domain, which provides discriminative contextual clues over the missing contents; and, 
	\item the \textit{Progressive Texture Blending} (PTB) module blends the encoded texture features to the main fusion stream in MEF in a multi-scale and progressive manner, and produces the final restoration results.
\end{itemize}

Throughout the system, we have incorporated several attention mechanisms to ensure the consistency of the fusion process, i.e., the \textit{Motion Attention} modules $\mathcal{F}_\text{mt}$, the \textit{Saturation Attention} modules $\mathcal{F}_\text{sat}$, and the \textit{Scale Attention} modules. The detail of these modules will be elaborated in the following subsections.


\subsection{The Multi-Exposure Fusion Module}

The input LDR image sequence is first transformed to the HDR domain with gamma correction and energy normalization according to:
\begin{equation}
H_k=\frac{I_k^{\gamma}}{\tau_k},~k\in\{s,~m,~l\}.
\end{equation}
The gamma correction process ($\gamma=2.2$) transforms the LDR images from a domain which is visually appealing to our eyes to a linear domain directly captured by the camera sensors \cite{sen2012robust}. Here $\tau_k$ indicates the respective exposure time for $I_k$, and the normalization brings all LDR images to the same \textit{exposure energy level}. 

The feature extraction module $\mathcal{F}_\text{ex}$ is applied over $\{H_s, H_m, H_l\}$ (with shared weights) to extract visual features $\{F_s, F_m, F_l\}$. To deal with the content discrepancy caused by camera motion and dynamic objects, the \textit{Motion Attention} modules $\mathcal{F}_\text{mt}^s$ and $\mathcal{F}_\text{mt}^l$ compare and detect the differences between the extracted features $F_s$, $F_l$ against $F_m$, and estimate the feature attention maps: $A_\text{mt}^{s}$ and $A_\text{mt}^{l}$. Any content misalignment in $F_s$, $F_l$ with respect to $F_m$ will be \textit{suppressed} by these attention maps. As illustrated in Fig.~\ref{fig:systemDiagram}(a), the \textit{Motion-suppressed Reference Exposure Features} can be formed and used as input to the MEF module by concatenating the features along the channel dimension: 
\begin{equation}
F_\text{mef}=[F_m, F_s \odot A_\text{mt}^s, F_l \odot A_\text{mt}^l].
\end{equation}
Here $\odot$ indicates point-wise multiplication.

The subsequent MEF module comprehensively explores the tonal profiles and signal correlations within $F_\text{mef}$. Specifically, a sequential concatenation of Channel Attention Blocks (CAB) \cite{zhang2018image} has been deployed to explore the channel-wise feature correlations. This helps to fully explore the characteristic information from different captures and regularize the signal distribution to the desired subspace. The MEF module determines the tonal mapping profile, suppresses noise, and enhances image details (contrast, sharpness, etc.).

\subsection{Progressive Neural Feature Transfer over Masked Saturated HDR Domain} \label{sec:domainTransform}

Shorter exposures in the capture sequence $\mathcal{I}$ reveal possible missing information in longer ones. The NFT module aims to transfer these missing information to the medium exposed image with accurate alignment against adversarial conditions such as camera motion and dynamic contents. 
The alignment process is challenging due to insufficient contextual clues, especially for larger saturation areas. Neural features provide powerful descriptions of signal correlations across multiple scales and imaging conditions, which prove to be efficient in cross-reference correspondence matching \cite{zhang2019image}. We propose a multi-scale Neural Texture Matching (NTM) mechanism to search for content correspondence in the Masked Saturated HDR (\textit{MS-HDR}) domain.

\begin{figure}[t]
	\centering
	\includegraphics[width=1\linewidth]{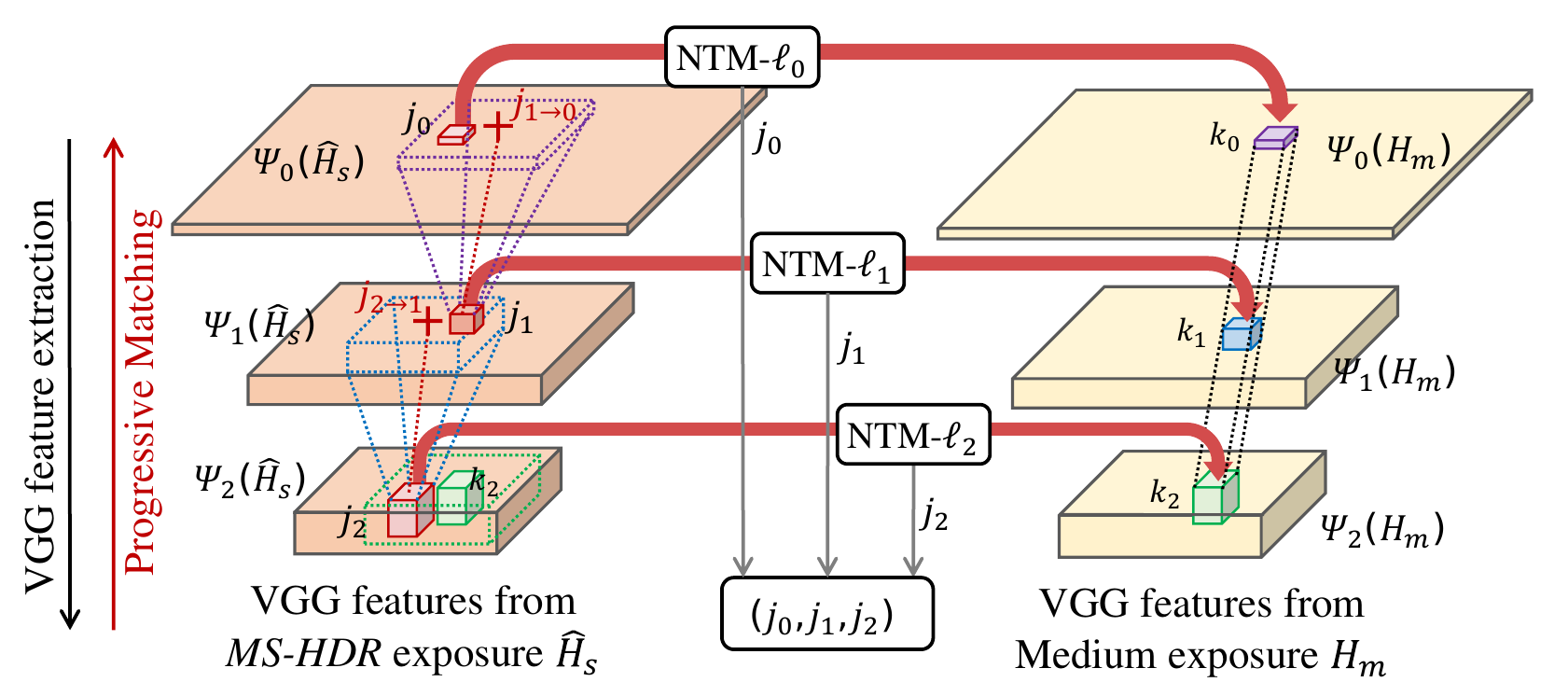}
	\caption{Multi-scale Progressive Neural Texture Matching (NTM) based on VGG features.}
	\label{fig:matchfeature}
\end{figure}

\subsubsection{\textbf{Masked Domain Transform}}

In order to promote signal similarity for efficient correspondence matching against saturation and motion (both camera and content motion), we transform the short-exposed HDR image $H_s$ into the artificial \textit{MS-HDR} domain $\hat{H}_s$ according to:
\begin{align*}
\hat{H}_s(x)=
\begin{cases}
\frac{\epsilon_m}{\tau_m} ,~ & \text{if}~H_s(x)>\frac{\tau_s}{\tau_m}\cdot \epsilon_s;\\
H_s(x),& \text{otherwise}.
\end{cases}
\end{align*}
Here $\epsilon_s$ is the saturation energy level for $I_s$, $\frac{\tau_s}{\tau_m}$ normalizes the saturation energy level from $H_m$ to $H_s$, making the well-exposed pixels in $H_s$ \textit{artificially saturated} similar to those in $H_m$.
It is assumed that after the transform, the saturated regions in $\hat{H}_s$ and $H_m$ will be identical, irrespective of foreground or camera motion; such assumption only fails for background\footnote{We refer to any region throughout the capture sequence as \textit{background}, as long as occlusion is present.} pixels with saturation. As illustrated in Fig. \ref{fig:domainTansform}, the histograms for $\hat{H}_s$ and $H_m$ become almost identical. The transform increases similarities between different exposures by actively masking out saturated textures; and the saturation ambiguity is expected to be resolve by associating surrounding textures. 

\subsubsection{\textbf{Progressive Neural Texture Matching}}
Based on $\hat{H}_s$ and $H_m$, we match the correspondence within a multi-scale neural feature pyramid.
The rationale for using a multi-scale framework is to involve contextual information outside of the saturated regions to anchor correspondence from a more global perspective. Same-sized patches at different scale levels will cover different content areas, which provides more comprehensive clues for robust feature matching. 

As illustrated in Fig. \ref{fig:matchfeature}, we denote the VGG feature extractor as $\Psi_l(H)$, which extracts multi-scale (scale indicated by the subscript $l$) features from $H$. We use $\mathcal{P}_i(\cdot)$ to denote sampling the $i$-th \textit{spatial} patch from the VGG feature maps.
Inner product $s_{i,j}$ is used to measure the feature similarity between the $i$-th \textit{MS-HDR} patch $\mathcal{P}_i(\Psi_l(\hat{H}_s))$ and the $j$-th HDR patch $\mathcal{P}_j(\Psi_l(H_m))$:
\begin{equation}\label{eq:similarity}
s_{i,j}=\Bigg \langle \mathcal{P}_i(\Psi_l(\hat{H}_s)), \frac{\mathcal{P}_j(\Psi_l(H_m))}{||\mathcal{P}_j(\Psi_l(H_m))||_2^2}\bigg \rangle.
\end{equation}
The similarity map computation can be efficiently implemented as convolution over $\Psi_l(\hat{H}_s)$ with $\mathcal{P}_i(\Psi_l(H_m))$ as the convolution kernel:
\begin{eqnarray}\label{eq:smap}
S^l_j= \Psi_l(\hat{H}_s) * \frac{\mathcal{P}_j(\Psi_l(H_m))}{||\mathcal{P}_j(\Psi_l(H_m))||_2^2},
\end{eqnarray}
where $*$ denotes the convolution operation. 

In order to promote cross-scale feature matching consistency and to reduce computation complexity, we adopt a progressive feature matching mechanism which restricts the calculation of the similarity map $S_j^l$ to a local window.
As illustrated in Fig. \ref{fig:matchfeature}, the progressive matching starts at the coarsest scale $l=2$. We use $S^l_j(k,\omega)$ to denote a \textit{local} similarity map within $\Psi_l(\hat{H}_s)$ at a neighborhood $\omega$ centered around pixel $k$. A best matched location $j_2$ within $\Psi_2(\hat{H}_s)$ can be found for the target patch $k_2$ from $\Psi_2(H_m)$ via:
\begin{equation}\label{eq:matching2}
j_2= \argmax_j S_j^2(k_2,\omega_2),
\end{equation}

For the next finer scale $l=1$, features will be matched within the local window $\omega_1$ centered around the pixel $j_{2\rightarrow 1}$, which is directly propagated from the lower level location $j_2$. A best match $j_1$ can be found via:
\begin{equation}\label{eq:matching1}
j_1= \argmax_j S_j^1(j_{2\rightarrow 1},\omega_1).
\end{equation}
Similarly, a best match $j_0$ can be found for the finest scale $l=0$. In the end, a tuple of best match locations $(j_0, j_1, j_2)$ at different VGG feature scales will be estimated for the corresponding target patch locations $(k_0, k_1, k_2)$.

\begin{figure}[t]
	\centering
	\includegraphics[width=1\linewidth]{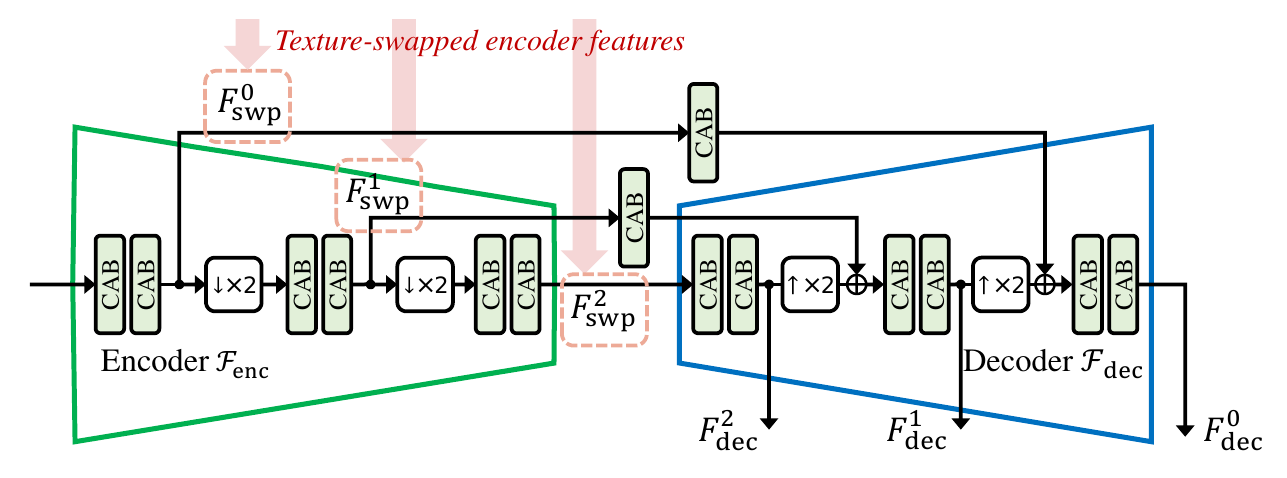}
	\caption{Structural details for the feature encoder/decoder modules $\mathcal{F}_\text{enc}$ and $\mathcal{F}_\text{dec}$. The encoded features $\{F^l_\text{swp}\}_{l=0}^2$ are from the NFT module, and the the decoded features $\{F^l_\text{dec}\}_{l=0}^2$ will be fused with features in the MEF module. The structure detail for the Channel Attention Block (CAB) is given in Fig. \ref{fig:systemDiagram}.}
	\label{fig:codec}
\end{figure}

\subsubsection{\textbf{VGG-Guided Neural Feature Transfer}}

The NFT module swaps the feature maps extracted by the Encoder $\mathcal{F}_\text{enc}$ to compensate for the missing contents caused by saturation. 
As illustrated in Fig. \ref{fig:systemDiagram}(b), the input to the feature encoder $\mathcal{F}_\text{enc}$ are the \textit{Motion-Suppressed Saturation Clue Features} which is formed by:
\begin{equation}
F_\text{nft}= F_s \odot \text{Sgm}[A_\text{mt}^s + A_\text{sat}^s],
\end{equation}
where $\text{Sgm}$ denotes the \textit{Sigmoid} function. $A_\text{sat}^s$ is the saturation attention predicted by the module $\mathcal{F}_\text{sat}$ based on the binary saturation mask of $\hat{H}_s$ in the \textit{MS-HDR} domain. $ A_\text{sat}^s$ helps to differentiate saturation from motion, and thus it encourages useful texture information to be transferred to regions with missing contents.

The structure details for $\mathcal{F}_\text{enc}$ is shown in Fig. \ref{fig:codec}. Similar to VGG, $\mathcal{F}_\text{enc}$ extracts visual features at three different scales, and each scale consists of two consecutive CAB blocks and a bilinear downsampler that reduces the feature spatial resolution by two.

Note that while the \textit{VGG features} $\Psi_l(\hat{H}_s)$ and $\Psi_l(H_m)$ are used for correspondence establishment as specified in Eq.~(\ref{eq:similarity}) to (\ref{eq:matching1}), the \textit{learned encoder features} $\Phi_l(F_\text{nft})$ and $\Phi_l(F_m)$ by $\mathcal{F}_\text{enc}$ are used for actual feature transfer. 
Based on the matching relationships $(k_0, k_1, k_2):\rightarrow(j_0, j_1, j_2)$, the patch $\mathcal{P}_k(\Phi_l(F_m))$ will be replaced by the corresponding features $\mathcal{P}_{j}(\Phi_l(F_\text{nft}))$. 
These swapped patched are finally formed into the \textit{texture-swapped encoder features} $\{F^l_\text{swp}\}_{l=0}^2$.

\textbf{Remark.} Using VGG features $\Psi_l(H_m)$ and $\Psi_l(\hat{H}_s)$ as matching guide proves to be important for identifying discriminative clues for robust matching against ambiguities caused by saturation. However, by actually swapping learned features $\Phi_l(F_m)$ and $\Phi_l(F_\text{nft})$, the network has a more consistent gradient flow for efficient feature learning and texture fusion. This will be validated in the ablation study in Sec. \ref{sec:ablation_study}.

\subsection{Progressive Texture Blending}

The decoder module $\mathcal{F}_\text{dec}$ takes the \textit{texture-swapped encoder features} $\{F_\text{swp}^l\}_{l=0}^2$ as input and outputs the decoder features $\{F_\text{dec}^l\}_{l=0}^2$. The structure of the decoder $\mathcal{F}_\text{dec}$ is illustrated in Fig. \ref{fig:codec}. It has similar structure with $\mathcal{F}_\text{enc}$ with skip connections from the encoder at each scale. 

To efficiently blend the decoder features $F^l_\text{dec}$ with the main MEF stream features in a \textit{tonal-} and \textit{contextual-}consistent manner, we introduce a progressive blending scheme where consistency is enforced via the \textit{Scale Attention} modules $\mathcal{F}_\text{sc}$ between different decoder scales. $\mathcal{F}_\text{sc}$ is made up of several fully convolutional layers with a \textit{Sigmoid} layer in the end, aiming at enforcing cross-scale consistency between different features scales. For the scale $l=1$, the scale attention map $A_\text{sc}^1$ will be estimated via:
\begin{equation}
A_\text{sc}^1= \mathcal{F}_\text{sc}^1[(F_\text{dec}^2)^{\uparrow_4}, (F_\text{dec}^1)^{\uparrow_2};~\theta_{sc}^1],
\end{equation}
where the superscript operator $[F]^{\uparrow_k}$ denotes scaling up the spatial resolution of $F$ by $k$-times via transposed convolution, and $\theta_\text{sc}^1$ denotes the model parameters to be learned. Similarly for the scale $l=0$, $A_\text{sc}^0$ will be estimated via:
\begin{equation}
A_\text{sc}^0= \mathcal{F}_\text{sc}^0[(F_\text{dec}^1)^{\uparrow_2}, F_\text{dec}^0; ~\theta_\text{sc}^0].
\end{equation}
For the coarsest scale $l=2$, the \textit{Scale Attention} map is directly set as the medium exposed image's saturation attention $A^m_\text{sat}$, which is predicted by the \textit{Saturation Attention} module $\mathcal{F}_\text{sat}$ based on the binary saturation map $M_\text{sat}^m$ from the medium exposure $H_m$: 
\begin{equation}
A^2_\text{sc}= A^m_\text{sat}= \mathcal{F}_\text{sat}[M_\text{sat}^m].
\end{equation}

The predicted attention maps will be multiplied with the features from corresponding scales and fused with the main fusion branch:
\begin{align}
H_\text{mef}= \mathcal{N}_\text{mef}[~A_\text{sc}^2 \odot (F_\text{dec}^2)^{\uparrow_4}, ~A_\text{sc}^1 \odot (F_\text{dec}^1)^{\uparrow_2}, & \notag\\~A_\text{sc}^0 \odot F_\text{dec}^0,~F_\text{mef}, ;~\theta_\text{mef}~&].
\end{align}

The final output of the APNT-Fuse model is calculated as residual and compensated to the medium capture modulated with weights $\mathcal{F}_\text{wgt}[A_\text{sat}^m]$:
\begin{equation}
H_\text{out} = H_m \odot (1- \mathcal{F}_\text{wgt}[A_\text{sat}^m]) + H_\text{mef}.
\end{equation}
Here, $\mathcal{F}_\text{wgt}$ denotes the Fusion Re-weighting module, which also consists of several fully convolutional layers with a \textit{Sigmoid} layer in the end. The final fusion weights between $H_m$ and $H_\text{mef}$ therefore depends on the \textit{Saturation Attention} $A_\text{sat}^m$. Note that the fusion weights are no longer binary but an optimized fusion ratio.


\subsection{Training Loss}

We focus on the visual quality of the fused HDR images after tone-mapping, therefore, we choose to train the network in the tone-mapped domain rather than the linear HDR domain. Given an HDR image $H_\text{out}$ in linear HDR domain, we compress the range of the image using the $\mu$-law \cite{kalantari2017deep}:
\begin{equation}\label{eq_mulaw}
\mathcal{T}(H_\text{out})= \frac{\text{log}(1+\mu H_\text{out})}{\text{log}(1+\mu)},
\end{equation}
where $\mu$ is a parameter defining the amount of compression, and $\mathcal{T}(H_\text{out})$ denotes the tone-mapped image. 
In this work, we always keep $H$ in the range $[0, 1]$ by adding a \textit{Sigmoid} layer at the end of the model, with $\mu$ set to 5000. The tone-mapper in Eq. (\ref{eq_mulaw}) is differentiable, which is most suitable for training the network.

We train the network by minimizing $\mathcal{L}_1$-norm based distance between the tone-mapped estimation $H$ and the ground truth HDR images $H_\text{gt}$:
\begin{equation}
\mathcal{L}= ||\mathcal{T}(H_\text{gt})- \mathcal{T}(H_\text{out})||_1.
\end{equation}

\subsection{Implementation Details}
We adopt a pre-trained VGG19 \cite{simonyan2015very} for feature extraction, which is well-known for its efficiency of texture representation \cite{gatys2015texture} \cite{gatys2016image}. Feature layers \textit{relu1\_1}, \textit{relu2\_1}, and \textit{relu3\_1} are used as the texture encoder. 
Adam optimizer \cite{kingma2014adam} was used for training, with batch size set to 1 and learning rate set to $1\times 10^{-5}$. We crop image into patches of size 256$\times$256 for training. Network weights are initialized using Xavier method~\cite{glorot2010understanding}.

\begin{figure*}[t]
	\centering
	\includegraphics[width= 0.98\linewidth]{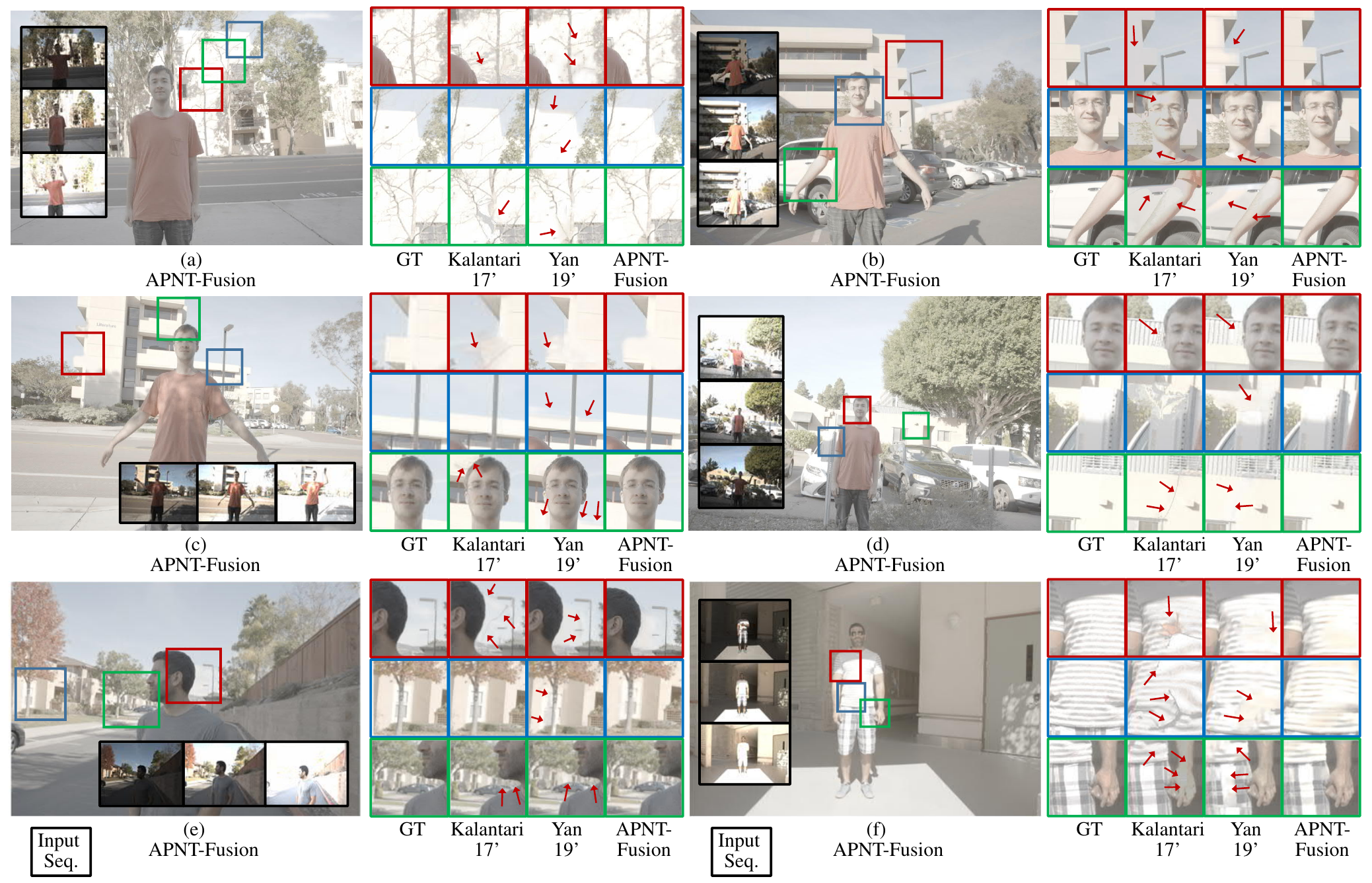}
	\caption{Visual Comparison for HDR restoration over dynamic scenes between Kalantari17' \cite{kalantari2017deep}, Yan19' \cite{yan2019attention}, and the proposed APNT-Fusion framework.}
	\label{fig:visualDemo}
\end{figure*}

\begin{figure*}[t]
	\centering
	\includegraphics[width= 0.98\linewidth]{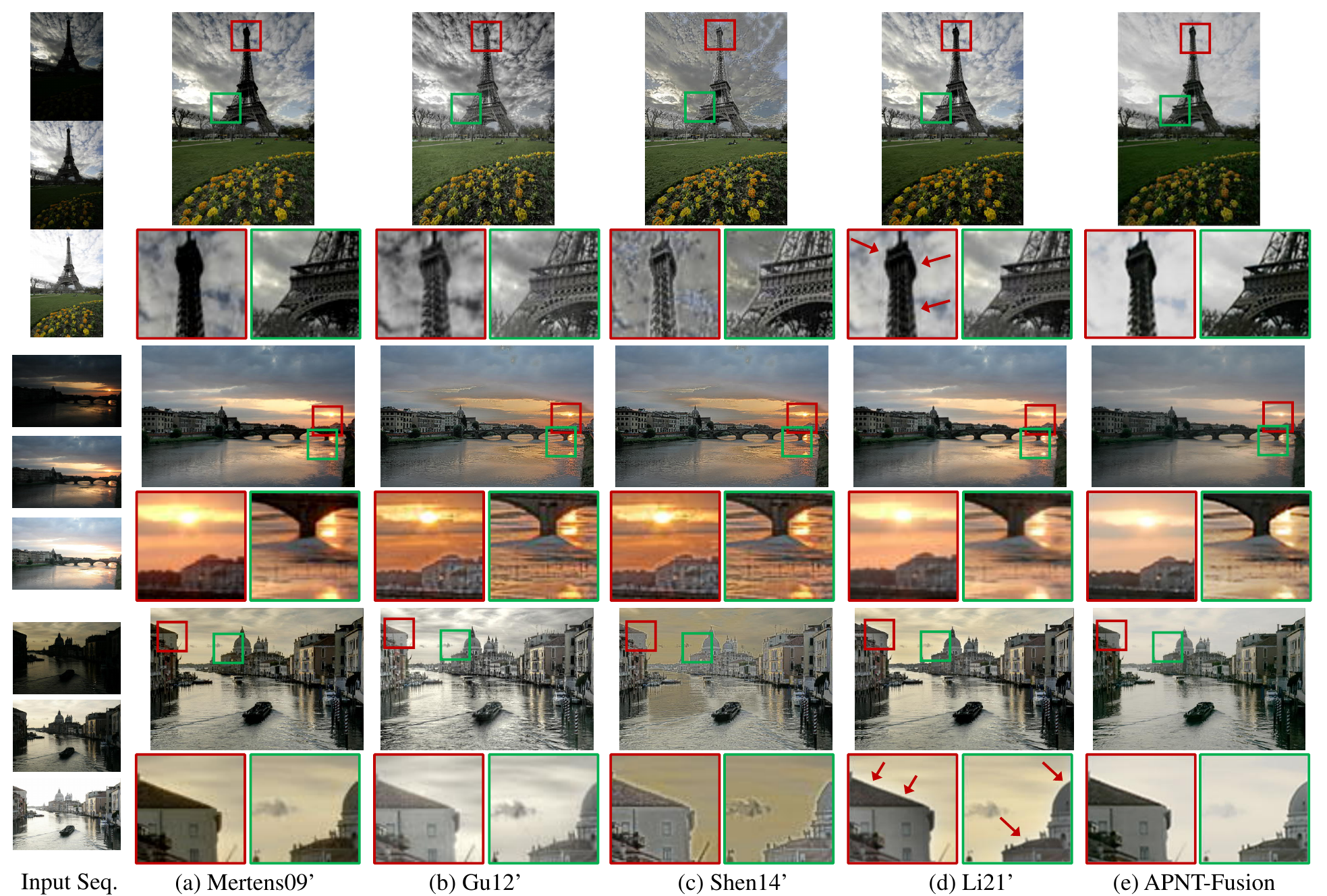}
	\caption{Visual Comparison for data from the MEF-Opt database between different methods: (a) Mertens09' \cite{mertens2009exposure}, (b) Gu12' \cite{gu2012gradient}, (c) Shen14' \cite{shen2014exposure}, (d) Li21' \cite{li2021detail}, and (e) our proposed APNT-Fusion framework.}
	\label{fig:staticCompare}
\end{figure*}

\begin{figure}[t]
	\centering
	\includegraphics[width=0.76\linewidth]{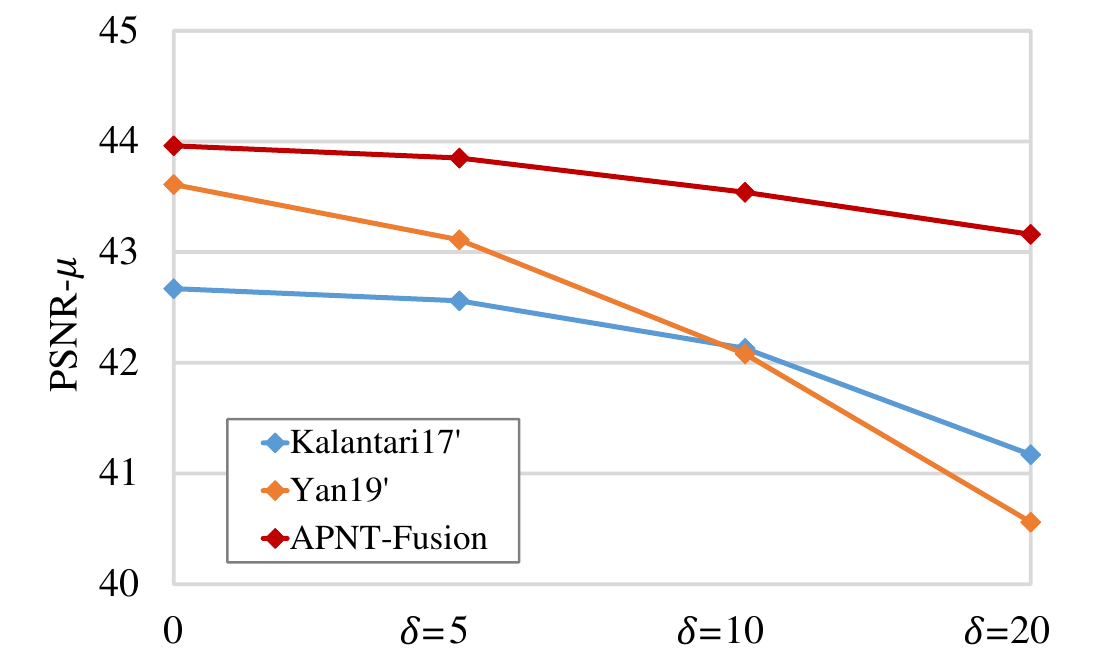}
	\caption{Comparison between Kalantari17' \cite{kalantari2017deep}, Yan19' \cite{yan2019attention} and APNT-Fusion on degradation of HDR restoration performance when translation (by $\delta$ pixels) is applied between the input LDR images.}
	\label{fig:shift}
\end{figure}\vspace{-0.4cm}

\setlength{\tabcolsep}{4pt}
\begin{table}
	\begin{center}
		\caption{Quantitative comparison of our proposed system against several state-of-the-art methods. The notation -$\mu$ and -L refer to the PSNR (in dB)/SSIM values calculated in the image tone-mapped (using Eq. (\ref{eq_mulaw})) and linear domains, respectively. The best and the second best results are highlighted in red and blue, respectively.} \label{table:psnr}
		\begin{tabular}{>{\arraybackslash}m{2.3cm} >{\arraybackslash}m{1.3cm} >{\arraybackslash}m{1.3cm} >{\arraybackslash}m{1.3cm} >{\arraybackslash}m{1.3cm}}
			\hline\noalign{\smallskip}
			& PSNR$-\mu$ &  PSNR$-L$  & SSIM-$\mu$ & SSIM-L\\
			\noalign{\smallskip}
			\hline
			\noalign{\smallskip}
			Wu et al. \cite{Wu2018deep} 		& 41.65		& 40.88 & 0.9860 & 0.9858\\
			Kalantari et al. \cite{kalantari2017deep} 	& 42.67  	& \textcolor{Blue}{\textbf{41.22}} & 0.9877 &0.9845\\
			Yan et al. \cite{yan2019attention} 	& \textcolor{Blue}{\textbf{43.61}}  	& 41.13 & \textcolor{Blue}{\textbf{0.9922}} &\textcolor{Blue}{\textbf{0.9896}}\\
			\hline
			APNT-Fusion		& \textcolor{Red}{\textbf{43.96}} 		& \textcolor{Red}{\textbf{41.69}} 	&\textcolor{Red}{\textbf{0.9957}} &\textcolor{Red}{\textbf{0.9914}}\\
			\hline
		\end{tabular}
	\end{center}
\end{table}
\setlength{\tabcolsep}{1.4pt}

\section{Experiments} \label{sec:experiments}

We comprehensively evaluate our model both quantitatively and qualitatively based on the benchmark HDR image datasets and compare our proposed method with state-of-the-art HDR restoration methods. The functionality of each component will be evaluated for their respective contributions in the ablation studies.

\subsection{Model Evaluation and Comparison}  \label{sec:evaluation}

\subsubsection{\textbf{The DeepHDR Dataset}} 
Proposed by Kalantari et al. \cite{kalantari2017deep}, the DeepHDR dataset includes bracketed exposures with dynamic contents: 74 groups of which are for training (each group contains 3 different exposures) and 15 groups for testing. All images are with resolution 1000$\times$1500 pixels. We conduct evaluations on different methods based on the following four metrics:
\textbf{PSNR}-L: the Peak Signal-to-Noise Ratio between the ground truth HDR image $H_\text{gt}$ and the direct output from the network $H_\text{out}$ in the linear HDR domain:
\begin{equation}
\text{PSNR-L}= -10\times \log_{10} \text{MSE}(H_\text{out}, H_\text{gt}),
\end{equation}
where $\text{MSE}$ represents Mean-Squared-Error for all pixels between $H_\text{gt}$ and  $H_\text{out}$. \textbf{PSNR}-$\mu$: the PSNR value between the ground truth and calculated HDR images in the tone-mapped domain based on the $\mu$-law in Eq. (\ref{eq_mulaw}). \textbf{SSIM}-L and \textbf{SSIM}-$\mu$: the Structural Similarity Index \cite{wang2004image} between the ground truths and calculated HDR images in the linear HDR and tone-mapped HDR domains, respectively. 

The quantitative results for the proposed APNT-Fusion are shown in TABLE~\ref{table:psnr}, which has achieved an average PSNR of \textbf{43.96} calculated across the RGB channels for all 15 testing images in the tone-mapped domain. We also compare with state-of-the-art methods, i.e., the 2-stage flow-based method \cite{kalantari2017deep} (denoted as Kalantari17'), the deep and fully convolutional restoration framework \cite{Wu2018deep} (denoted as Wu18'), and the attention guided HDR framework \cite{yan2019attention} (denoted as Yan19'). The results are shown in Table~\ref{table:psnr}. As can be seen, our model APNT-Fusion achieves the highest PSNR and SSIM values in both linear and tone-mapped domains. Although the quantitative advantage of APNT-Fusion is around \textbf{\textit{0.35 dB}} against Yan19' in PSNR-$\mu$, we believe it is caused by the relatively small saturation area in the testing dataset. In the following visual comparison, we show that the advantage of APNT-Fusion is very obvious against all other methods.

We carry out qualitative comparison, and the results are shown in Fig.~\ref{fig:visualDemo}. We focus on the most challenging areas, i.e., the restoration of saturated regions, and the details are highlighted in the zoom-in boxes. As can be seen, outputs from Kalantari17' introduce undesirable artifacts in ambiguous motion regions. The introduction of advanced optical flow regularization solves the ambiguity issues associated with texture-less saturated regions to some extent, but large areas of saturated pixels remain in the output image as shown in Fig.~\ref{fig:visualDemo}(b) and (d). In addition, obvious distortions can be observed in Fig.~\ref{fig:visualDemo}(c)-(f). With the VGG-guided matching mechanism, APNT-Fusion estimates correspondence more accurately for ambiguous regions, especially the challenging areas for large saturated regions.

The attention-based network Yan19' handles motion boundaries much better than Kalantari17'. Nevertheless, the disadvantage is also obvious. While the attention mask suppresses pixel discrepancies between different reference images which reduces ghosting artifacts, it also suppresses the transfer of useful information to the saturated pixels. Such artifacts are especially obvious over the silhouette between the building and the bright sky in Fig. \ref{fig:visualDemo}(a), (b) and (c). The boundaries are faded since no mechanism has been designed to distinguish between saturation and motion. Due to the introduction of motion attention and the multi-scale progressive fusion mechanism, APNT-Fusion show much better restoration results in texture transfer to saturated regions and in the preservation of well-exposed structures.

\textbf{\textit{Robustness against camera motion.}} We carry out experiments on the robustness of each method when camera motion is artificially imposed. We impose horizontal and vertical translation of $\delta=\{0,~5,~10~,20\}$ pixels for $I_l$ and $I_s$ with respect to $I_m$ over all sequences of the testing dataset. The HDR restoration results are shown in Fig.~\ref{fig:shift} with PSNR-$\mu$ values being the vertical axis and $\delta$ values being the horizontal axis. As can be seen, the performance for Yan19' deteriorates quickly as no pixel association mechanisms are employed in their framework. The misalignment causes the motion attention to suppress most of the valuable information, leading to a fast performance decay. Although optical flow was employed for Kalantari17', we can still observe a sharper performance decline as $\delta$ becomes larger compared with APNT-Fusion. This experiment validates APNT-Fusion's robustness against motion due to the multi-scale neural feature matching mechanism.

\begin{figure}[t]
	\centering
	\includegraphics[width= 1\linewidth]{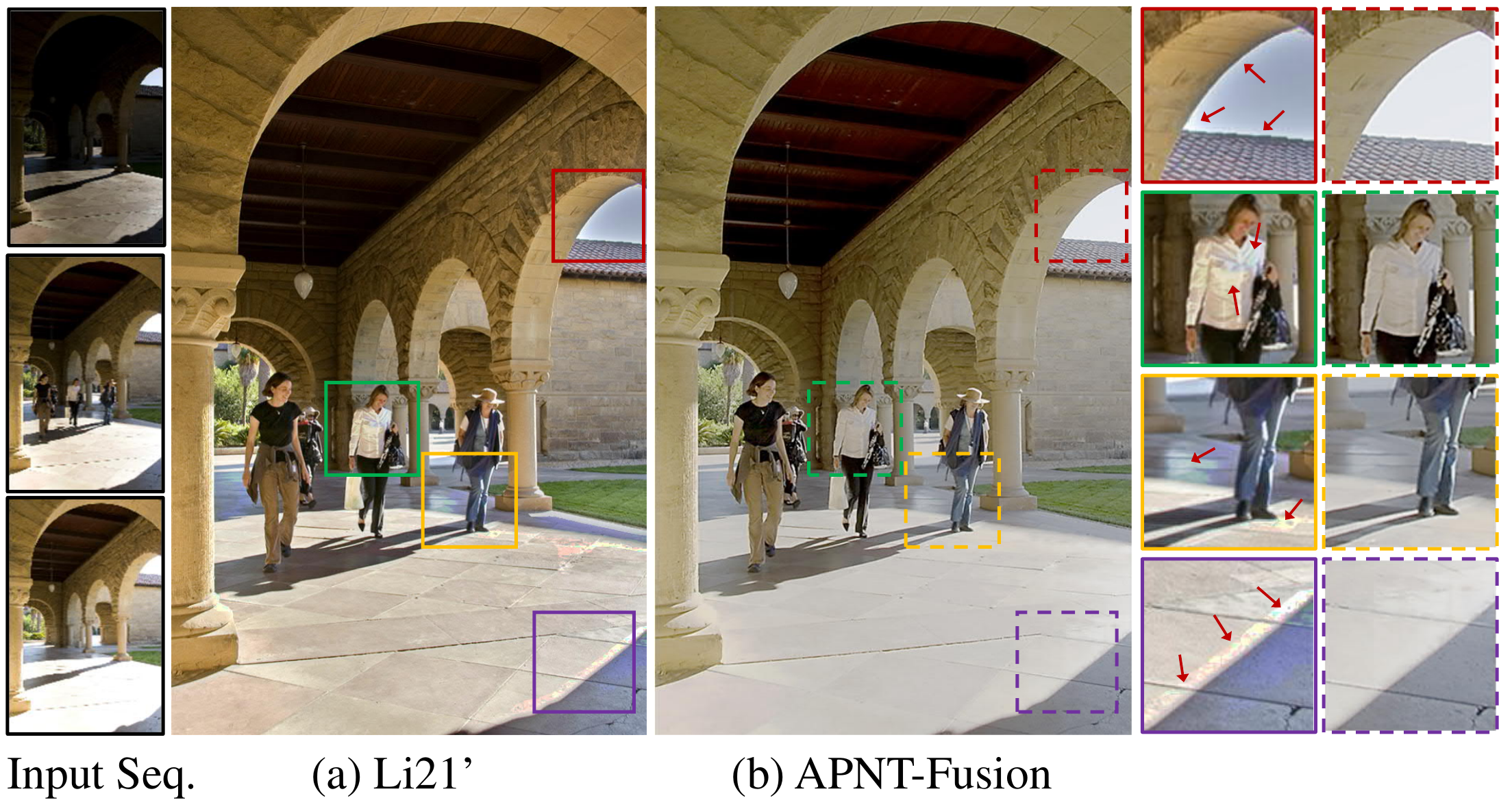}
	\caption{Visual Comparison for dynamic contents between (a) Li21' \cite{li2021detail} and (b) the proposed APNT-Fusion.}
	\label{fig:dynamicCompare}
\end{figure}

\setlength{\tabcolsep}{4pt}
\begin{table}
	\begin{center}
		\caption{Quantitative ablation study of our proposed APNT-Fusion model against several variant networks. Most significant factors highlighted in red and blue, respectively.} \label{table:ablation}
		\begin{tabular}{>{\arraybackslash}m{2.3cm} >{\arraybackslash}m{2.3cm} >{\arraybackslash}m{2.3cm}}
			\hline\noalign{\smallskip}
			& PSNR$-\mu$  & SSIM-$\mu$ \\
			\noalign{\smallskip}
			\hline
			\noalign{\smallskip}		
			w/o \textit{MS-HDR}& 43.11 (-0.85) &0.9869 (-0.0088)\\
			w/o NFT	& 42.37 (\textcolor{Blue}{\textbf{-1.59}}) &0.9831 (\textcolor{Blue}{\textbf{-0.0126}})\\
			\hline
			w/o VGG-$\mathcal{L}_2\mathcal{L}_3$	& 43.68	(-0.28) &0.9929 (-0.0028)\\
			w/o VGG w $\mathcal{F}_\text{enc}$ & 43.57	(-0.39) &0.9927 (-0.0030)\\
			\hline
			w/o Motion Att.	& 41.73	(\textcolor{Red}{\textbf{-2.23}})	 &0.9814 (\textcolor{Red}{\textbf{-0.0143}})\\
			w/o Scale Att.	& 43.35	(-0.61) &0.9934 (-0.0023)\\
			\hline
			APNT-Fusion		&43.96  &0.9957 \\
			\hline
		\end{tabular}
	\end{center}
\end{table}
\setlength{\tabcolsep}{1.4pt}

\subsubsection{\textbf{The MEF-Opt Database}} Proposed by Ma et al. \cite{ma2017multi}, the MEF-Opt database contains 32 sets of multi-exposure image sequences, most of which are static scenes without well-exposed HDR ground truths for direct quantitative evaluation. 
For visual comparison, the results for different static multi-exposure fusion methods are shown in Fig. \ref{fig:staticCompare}. As can be seen, the proposed APNT-Fusion framework generally produces better fusion outcomes against the other state-of-the-art HDR fusion methods. Our results are with clearer boundaries between bright and dark regions. The halo effect has been much better suppressed due to the deep regularization of the MEF module. The textures have been well fused to the over-exposed regions, as highlighted in the zoom-in boxes.

We show the visual comparison for scenarios with dynamic objects between APNT-Fusion and the method proposed by Li et al. \cite{li2021detail} (denoted as Li21') in Fig. \ref{fig:dynamicCompare}. As can be seen, APNT-Fusion still consistently shows advantages in suppressing halo effects and restoring details over regions with both saturation and motion.

Note that the MEF-Opt dataset was built for multi-exposure image fusion in the image domain (with 8-bit unsigned integers as data format), and there are no ground truth HDR images in the radiance domain available for direct quantitative evaluation. Although metrics such as the MEF-SSIM score \cite{ma2017multi} has been proposed as quantitative measures on the fusion quality, it is unfair to use such metrics to compare with methods that work in different domains, as the domain shift affects the MEF-SSIM scores significantly without truthfully reflecting the reconstruction visual quality. Therefore, we do not compare MEF-SSIM in this experiment.

\subsection{Ablation Study} \label{sec:ablation_study}

To comprehensively evaluate separate modules of our framework, we carried out the following ablation studies. Specifically, we are going to independently evaluate the contributions of the texture transfer module and the attention fusion modules. Note that all the networks have been independently trained from scratch with the same training data and training settings as the complete APNT-Fuse model. The results shown in TABLE~\ref{table:ablation} are testing outcomes based on the 15 testing images from the DeepHDR \cite{kalantari2017deep} dataset.

\subsubsection{\textbf{Contribution of the Neural Feature Transfer Module}}
To evaluate the contribution of the NFT module, the following two derivatives of networks are designed for performance analysis:
\begin{itemize}
\item \textbf{w/o \textit{MS-HDR}}: The VGG features of $\Psi(H_m)$ are no longer matched with $\Psi(\hat{H}_s)$ in the \textit{MS-HDR} domain, but instead, directly matched with $\Psi(H_s)$. As shown in TABLE~\ref{table:ablation}, the transformation of $H_s$ from the HDR domain to the \textit{MS-HDR} domain during VGG correspondence matching brings performance advantage of around \textbf{\textit{0.83~dB}}. This proves our claim that more accurate matching can be achieved in the \textit{MS-HDR} domain.
\item \textbf{w/o NFT}: No neural feature transfer is implemented. The encoded features of $\mathcal{F}_\text{enc}(F_m)$ are directly used for Progressive Texture Blending. When the entire neural transfer module is removed, the performance of the network \textbf{HDR w/o NT} dropped by \textbf{\textit{1.59 dB}}, which signifies the contribution of the \textit{Neural Texture Transfer Network}.
\end{itemize}


	

\begin{figure}[t]
	\centering
	\includegraphics[width=1\linewidth]{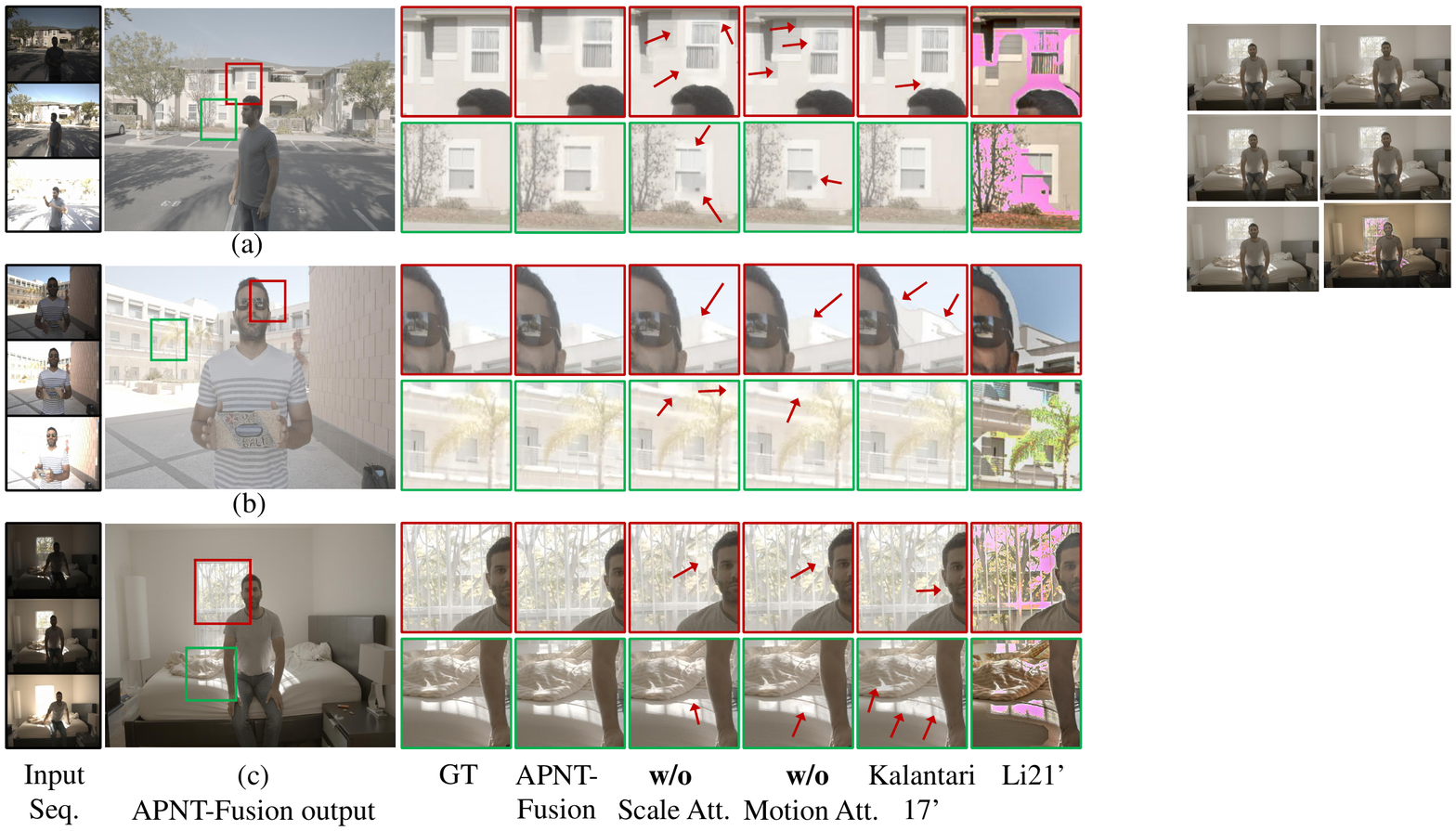}
	\caption{Visual Comparison for ablation study over the full model, w/o scale attention, and w/o motion attention. Results from Kalantari17' \cite{kalantari2017deep}, and Li21' \cite{li2021detail} are also shown for comparison.}
	\label{fig:ablation}
\end{figure}


\subsubsection{\textbf{Contribution of the VGG Feature Matching Module}}
In our framework, we have adopted the mechanism to search for neural feature correspondence based on multi-scale VGG neural features. Based on the matching outcomes, actual neural features are swapped in the encoded feature space subsequently. We carry out ablation study for this mechanism based on the following two setups:
\begin{itemize}
\item \textbf{w/o VGG-$\mathcal{L}_2\mathcal{L}_3$}: in this setup, we test only using the original scale of the VGG features \textit{relu1\_1} for correspondence matching. The results in Table \ref{table:ablation} shows that a \textbf{\textit{0.28 dB}} advantage is achieved by the multi-scale mechanism. Such a multi-scale scheme is useful in resolving ambiguities for larger saturated regions. 
\item \textbf{w/o VGG w $\mathcal{F}_\text{enc}$}: in this setup,  we test not using VGG features for matching, but instead, we directly rely on learned encoder features in $\mathcal{F}_\text{enc}$ for \textit{\ul{both feature matching and swapping}}. As can be seen, a performance drop by \textbf{\textit{0.39 dB}} is observed, which supports our claim that VGG features provides more discriminative clues for establishing correspondence against ambiguities of various sorts.
\end{itemize}

\subsubsection{\textbf{Contribution of the Attention Fusion Network}}

\begin{itemize}
\item \textbf{w/o Motion Att.}: in this setup, we test the contribution of the motion attention module $\mathcal{F}_\text{mt}$ by setting all elements in the motion attention maps $A^s_\text{mt}$ and $A^l_\text{mt}$ to be 1. This means the features from all exposures are directly concatenated as $[F_m, F_s, F_l]$ and fed to the MEF module for fusion. As can be seen, a drop of around \textbf{\textit{2.23 dB}} is observed when the Motion Attention module is absent. This consolidates the contribution of such mechanism in preventing ghosting artifacts.
\item \textbf{w/o Scale Att.}: We test the contribution of scale attention module $\mathcal{F}_\text{sc}$ by setting all elements in the scale attention maps $A^0_\text{sc}$ and $A^1_\text{mt}$ to be 1. Note that $A^2_\text{sc}$ is still set to be equal to $A^m_\text{sat}$, which contains no cross-scale information. In TABLE \ref{table:ablation}, a drop of \textbf{\textit{0.61 dB}} is observed when the scale attention module is removed. This validates the effectiveness of this module for preserving consistency when progressively blend transferred textures to the multi-exposure fusion stream.
\end{itemize}

Visual comparisons for the ablation studies on the attention modules are shown in Fig. \ref{fig:ablation}. As can be seen, without the scale attention modules, larger saturated regions show inconsistent texture fusion; however, with cross-scale consistency enforced, the texture transfer is much more reliable. In addition, we can also see that without motion attention modules, content misalignment caused obvious distortions after exposure fusion. We have also put Kalantari17' \cite{kalantari2017deep} and Li21' \cite{li2021detail} in Fig. \ref{fig:ablation} for visual comparison. It's worth mentioning that obvious color distortions can be observed from the results of Li21'; this is because the bright region has been misclassified as motion area, over which histogram equalization is applied to restore contrast, causing the unpleasant artifacts. 

Through the ablation studies, we have validated the important roles the novel modules play in the AFNT-Fusion framework.


\section{Conclusion} \label{sec:conclusion}

In this work, we have proposed an Attention-guided Progressive Neural Texture Fusion (APNT-Fusion) HDR restoration framework, which addresses the issues of motion-induced ghosting artifacts prevention and texture transfer over saturated regions efficiently within the same framework. 
A multi-scale Neural Feature Transfer module has been proposed to search for content correspondence via masked saturated transform, which actively masks out saturated textures and associates surrounding textures to resolve ambiguity. Transferred neural features are then combined to predict the missing contents of saturated regions in a multi-scale progressive manner with novel attention mechanisms to enforce cross-scale tonal and texture consistency. 
Both qualitative and quantitative evaluations validate the advantage of our method against the state-of-the-art solutions.

\bibliographystyle{IEEEbib}
\bibliography{mybib}

\end{document}